\documentclass[prd,preprint,tightenlines,floatfix,showpacs,preprintnumbers,nofootinbib,eqsecnum,superscriptaddress]{revtex4}
\usepackage[cp1250]{inputenc} 
\usepackage[T1]{fontenc} 
\usepackage{amsfonts} 
\usepackage{amsopn}   
\usepackage{amssymb}  
\usepackage{amsthm}   
\usepackage{graphicx}
\usepackage{mathtools}
\usepackage{cancel}
\usepackage{enumerate}
\newcommand{\bq}{\mbox{\boldmath $q$}}
\newcommand{\br}{\mbox{\boldmath $r$}}

\newcommand{\bs}{\mbox{\boldmath $s$}}

\newcommand{\bDelta}{\mbox{\boldmath $\Delta$}}

\newcommand{\bc}{\mbox{\boldmath $c$}}
\newcommand{\bb}{\mbox{\boldmath $b$}}
\newcommand{\bB}{\mbox{\boldmath $B$}}

\newcommand{\ket}[1]{| {#1} \rangle}
\newcommand{\bra}[1]{\langle {#1} |}

\newcommand{\half}{{1\over 2}}
\def\Pom{{\bf I\!P}}
\def\lsim{\mathrel{\rlap{\lower4pt\hbox{\hskip1pt$\sim$}}
		\raise1pt\hbox{$<$}}}         
\def\gsim{\mathrel{\rlap{\lower4pt\hbox{\hskip1pt$\sim$}}
		\raise1pt\hbox{$>$}}}         

\begin{document}

\vfill
\title{Incoherent diffractive photoproduction of $J/\psi$ and $\Upsilon$ on heavy nuclei in the color dipole approach}

\author{Agnieszka {\L}uszczak}
\email{agnieszka.luszczak@desy.de} 
\affiliation{
	T.~Kosciuszko Cracow University of Technology, PL-30-067 
	Cracow, Poland}
\affiliation{Deutsches Elektronen-Synchrotron DESY, D-22607 Hamburg,
	Germany}

\author{Wolfgang Sch\"afer}
\email{Wolfgang.Schafer@ifj.edu.pl} \affiliation{Institute of Nuclear Physics Polish Academy of Sciences, ul. Radzikowskiego 152, PL-31-342 Cracow, Poland}

\date{\today}

\begin{abstract}
We calculate cross sections and transverse momentum distributions for the incoherent diffractive
production of vector mesons $J/\psi$ and $\Upsilon$ on heavy nuclei. In distinction to coherent diffraction, 
the nucleus is allowed to break up, but except for the  vector meson no new particles 
are produced in the reaction. 
Within the color dipole approach, we derive the multiple scattering expansion of the incoherent 
diffractive cross section as an expansion over quasielastic scatterings of the color dipole.  
We also compare our results to the measurement of the ALICE collaboration for incoherent $J/\psi$
production at $\sqrt{s_{\rm NN}} = 2.76 \, \rm {TeV}$ and show predictions 
at $\sqrt{s_{\rm NN}} = 5.02 \, \rm{TeV}$. We also briefly discuss a possible contribution
to $J/\psi$ production in peripheral collisions in the $70 \div 90 \%$ centrality class.	
\end{abstract}

\pacs{13.87.-a, 11.80La,12.38.Bx, 13.85.-t}


\maketitle

\section{Introduction}

There has been recently a renewed interest in the diffractive photoproduction
of vector mesons on heavy nuclei, especially in connection with ultraperipheral heavy-ion
collision at RHIC and the LHC, see for example 
the reviews \cite{Contreras:2015dqa,Bertulani:2005ru,Baur:2001jj}.
Photoproduction of vector mesons on nuclei has been studied for a long time, 
mainly as a probe of the hadronic structure of the photon. 
The early works, mainly on photoproduction of light vector mesons are formulated within the
vector meson dominance model (see ref. \cite{Bauer:1977iq} for a review).

Nuclear effects in diffractive photoproduction are then treated as a vector-meson nucleus scattering 
following the rules of Glauber theory \cite{Glauber}, developed for hadron nucleus scattering 
at high energy. The energy should be not too high, though: at some point diffractive 
dissociation of hadrons becomes important and Glauber theory needs to be amended introducing the so-called 
inelastic shadowing corrections \cite{Wilkin}, a review can be found in \cite{Alberi_Goggi}.

\begin{figure}[!h]
	\begin{center}
		\includegraphics[width=.3\textwidth,angle = 270]{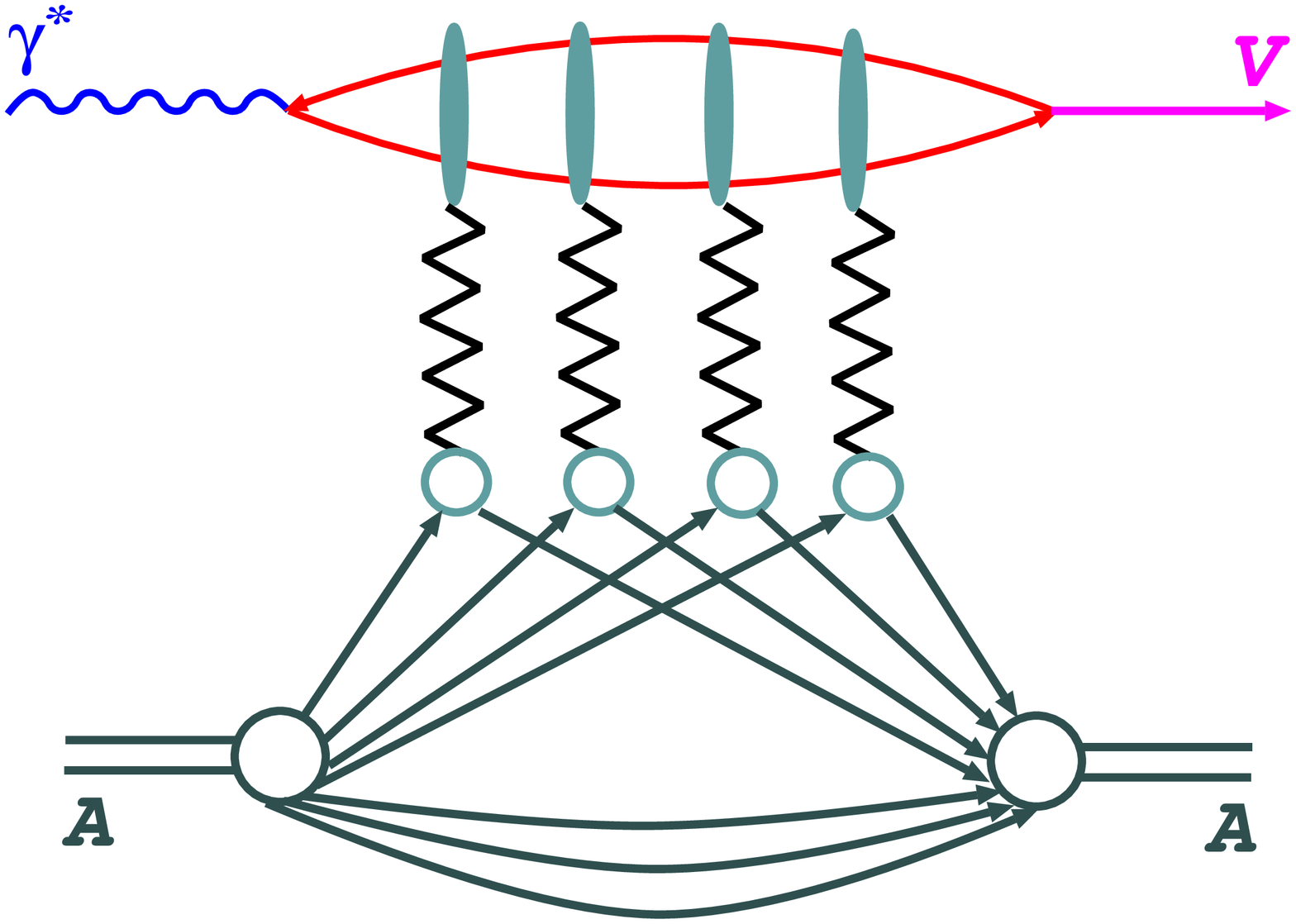}
		\includegraphics[width=.3\textwidth,angle = 270]{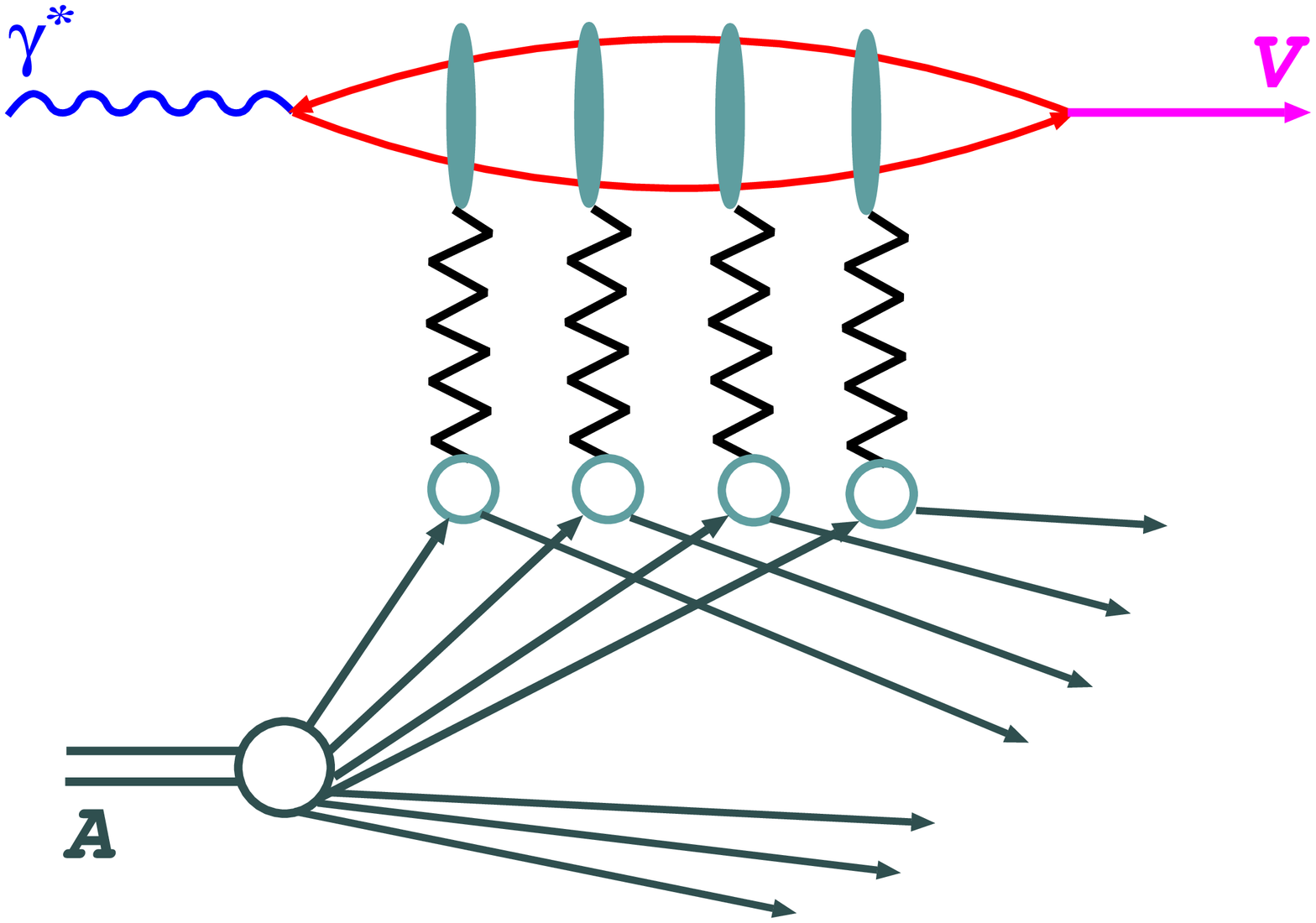}
		\caption{Left: coherent photoproduction of a vector meson in which the nucleus stays in its ground state. 
			Right: an example diagram for incoherent diffractive production, in which the nucleus breaks up. 
		}
		\label{fig:diagrams}
	\end{center}
\end{figure}

Here we discuss a specific inelastic reaction, the incoherent diffractive photoproduction.
In distinction to coherent diffraction (see the left diagram of fig. \ref{fig:diagrams}), 
here the target nucleus breaks up (right diagram of fig. \ref{fig:diagrams}). There is a large
rapidity gap between the produced vector meson and the nuclear fragments -- the signature of 
any diffractive mechanism. The distinction from generic inelastic diffraction is that there
are no new particles produced in the nuclear fragmentation region.

The nuclear final state may contain discrete excited states of the target, but in general will consist
of a continuum of fragments of a variety of charge to mass ratios, among them free protons and neutrons.
The description of the nuclear fragmentation region requires quite involved modeling of the nuclear 
dynamics, for a discussion of related problems, see e.g. \cite{Battistoni:2007zzb}. 

The problem becomes tractable, if we restrict ourselves to the sum over all possible nuclear 
states \cite{Glauber}. In hadron-nucleus scattering the corresponding theory of incoherent or
quasielastic processes was developed in \cite{Czyz:1970sq,Glauber:1970jm}. 
The single-channel formalism found there applies in the energy range where diffractive dissociation
is not yet important. The corresponding multi-channel generalization including Gribov's inelastic
shadowing corrections was given in \cite{Kolya_DSE}. 

In this work we are interested in the production of vector mesons built of heavy quarks 
$Q \bar Q$ ($Q$ = charm or bottom) in the 
high energy limit, where the formation time of the vector meson exceeds the nuclear size. 
In this case the appropriate frame work is the color dipole approach, where $Q \bar Q$ states
of fixed transverse size $\br$ interact with the nucleus and play the role of eigenstates
of the diffractive $S$-matrix \cite{Nikolaev:1992si}.

Recent works on coherent and incoherent vector meson production in the color dipole approach are
\cite{Goncalves:2011vf,Goncalves:2015poa,Ducati:2017bzk,Goncalves:2017wgg}, which are based on a similar Glauber-Gribov approach as in
our work. In ref. \cite{Ivanov:2004dg} the relation of nuclear attenuation to the so-called saturation
scale is discussed, and the behaviour of helicity-flip observables in the strong attenuation regime 
is derived. 
The formalism of \cite{Gustafson:2015ada}  also employs the color dipole approach and lends itself
to the calculation of incoherent diffractive observables, it is geared towards applications
also for soft processes. The hot-spot models \cite{Mantysaari:2017dwh,Cepila:2017nef} 
are also based on the color dipole approach, but use a different way of
averaging over target states. 
The treatment of coherent photoproduction in $k_\perp$-factorization in \cite{Cisek:2012yt} is equivalent
to the color dipole approach, however the calculation of incoherent diffraction
becomes unnecessarily cumbersome in this formulation.
An approach implementing the Glauber-Gribov shadowing not
based on color dipoles is found in \cite{Guzey:2013jaa}.   

\section{Incoherent photoproduction in the color dipole approach}
\label{sec:formalism}

We start by writing the amplitude for the reaction $\gamma A_i \to V(\bDelta) A_f^*$ for a finite transverse momentum
$\bDelta$ carried by the vector meson, as
\begin{eqnarray}
\label{wzor1}
\mathcal{A}(\gamma^* A_i \rightarrow VA^*_f;W,\Delta) &=& 
2 i \, \int d^2\bB \exp[-i\bDelta \bB] \bra{V} \bra{A^*_f} \hat \Gamma(\bb_+, \bb_- ) \ket{A_i}\ket{\gamma} \nonumber \\
&=& 2 i \, \int d^2\bB \exp[-i\bDelta \bB] \nonumber \\ 
&\times& \int_0^1 dz \int d^2\br 
\Psi_V^*(z,\br) \Psi_\gamma(z,\br) \bra{A^*_f} \hat \Gamma (\bB-(1-z)\br , \bB + z \br) \ket{A_i} . \nonumber \\
\end{eqnarray}
In order not to clutter the notation, we suppressed the sum over quark and antiquark helicities as well as the dependence
of $\hat \Gamma$ on center-of-mass energy (per nucleon) $W$.
Above $\bB$ is the impact parameter of the vector meson (or incoming photon), which is the proper conjugate variable to the
vector meson transverse momentum. Notice that the impact parameters of quark $\bb_+$ and antiquark $\bb_-$ which share 
the incoming photon (or outgoing vector meson) longitudinal momentum in fractions $z$ and $1-z$.
The equality
\begin{eqnarray}
\bB = z \bb_+ + (1-z) \bb_- \, ,
\end{eqnarray}
can be understood as a conservation of orbital angular momentum. Equation \ref{wzor1} has the disadvantage that it couples the 
$z$, $\bB$ and $\br$ integrations in a complicated manner.
It is therefore more convenient to go to new variables
\begin{eqnarray}
\bb = {\bb_+ + \bb_- \over 2} ,  \, \br = \bb_+ - \bb_- .
\end{eqnarray}
Because of 
\begin{eqnarray}
\bB = \bb - (1-2z) {\br \over 2} \, ,
\end{eqnarray}
the amplitude (\ref{wzor1}) now takes the form
\begin{eqnarray}
\mathcal{A}(\gamma^* A_i \rightarrow VA^*_f;W,\Delta) &=& 
2i \int d^2\bb \exp[-i \bb\bDelta] \int d^2\br \rho_{V\gamma}(\br,\bDelta)  \bra{A^*_f} \hat \Gamma (\bb + {\br \over 2} , \bb - {\br \over 2}) \ket{A_i} \, , \nonumber \\
\end{eqnarray}
where
\begin{eqnarray}
\rho_{V \gamma}(\br,\bDelta) = \int_0^1 dz \exp[i(1-2z) {\br \bDelta \over 2}] \Psi_V^*(z,\br) \Psi_\gamma(z,\br) \, , 
\label{eq:overlap}
\end{eqnarray}
which gives us an additional $\bDelta$--dependence, not coming from the Fourier transform of the nuclear amplitude.
Notice however, that the lightcone wave function of heavy vector mesons is sharply peaked around $z \sim 1/2$, and the
phase factor in eq. \ref{eq:overlap} can be safely neglected. It is indeed a relativistic effect.
By the same token, the variable $\bb$, which is just a dummy variable without direct physical meaning, becomes 
equal to the physical impact parameter at $z \to 1/2$. 

We turn to the derivation of the relevant differential cross section.
Our amplitude is normalized, so that
\begin{eqnarray}
{d \sigma (\gamma A_i \to V A_f^*)\over d\bDelta^2} = { 1 \over 16 \pi} \Big| \mathcal{A}(\gamma^* A_i \rightarrow VA^*_f;W,\Delta) \Big|^2 \, , 
\end{eqnarray}
The incoherent cross section of interest is defined as
\begin{eqnarray}
{d\sigma_{\rm incoh} \over d\bDelta^2} = \sum_{A_f \neq A} {d \sigma (\gamma A_i \to V A_f^*)\over d\bDelta^2} \, .
\end{eqnarray}
Using completeness in the sum over nuclear final states (which will include continuum states as well as bound states)
\begin{eqnarray}
\sum_{A\neq A_f}|A_f\rangle\langle A_f|=1-|A\rangle\langle A|,
\end{eqnarray}
we obtain the differential cross section in the form
\begin{eqnarray}
{d \sigma_{\rm incoh} \over d\bDelta^2} = { 1 \over 4 \pi} \int d^2\br d^2\br' 
\rho^*_{V\gamma}(\br',\bDelta) \rho_{V \gamma}(\br,\bDelta) \Sigma_{\rm incoh} (\br, \br',\bDelta) \, ,
\label{eq;dsig_dt}
\end{eqnarray}
with 
\begin{eqnarray}
&&\Sigma_{\rm incoh}(\br,\br',\bDelta) = \int d^2\bb d^2\bb' \exp[-i \bDelta(\bb - \bb')] 
\mathcal{C} \Big(\bb'+{\br' \over 2}, \bb' -{\br' \over 2}; \bb+{\br \over 2}, \bb -{\br \over 2} \Big) 
\label{eq:Sigma}
\end{eqnarray}
Here the function $\mathcal{C}$ involves only the ground-state nuclear averages
\begin{eqnarray}
\mathcal{C}(\bb'_+,\bb'_-;\bb_+,\bb_-) = \bra{A} \hat \Gamma^\dagger (\bb'_+, \bb'_-)\hat \Gamma(\bb_+, \bb_-) \ket{A}
- \bra{A} \hat \Gamma(\bb'_+, \bb'_-) \ket{A}^* \bra{A} \hat \Gamma(\bb_+, \bb_-) \ket{A} \, .
\end{eqnarray}
In performing these nuclear averages we will follow standard prescriptions of Glauber theory \cite{Glauber}, 
where the nuclear profile function operator is written as
\begin{eqnarray}
\hat \Gamma(\bb_+,\bb_-) = 1 - \prod_{i=1}^A [ 1 - \hat \Gamma_{N_i}(\bb_+ -\bc_i, \bb_--\bc_i)] \, ,
\end{eqnarray}
and we will accepting the standard dilute gas approximation of uncorrelated nucleons.

The profile functions for the free nucleon are related to the off-forward (finite transverse momentum transfer) dipole 
amplitude \cite{Nemchik:1997xb} as
\begin{eqnarray}
\sigma(\br,\bq) = 2 \int d^2\bb \exp[i \bq \bb] \bra{N} 
\hat \Gamma_N(\bb+{\br \over 2}, \bb -{\br \over 2}) \ket{N} \equiv  2 \int d^2\bb \exp[i \bq \bb] \hat \Gamma_N(\bb+{\br \over 2}, \bb -{\br \over 2}) \, .
\nonumber \\
\label{eq:dipole_amplitude}
\end{eqnarray}
The coherent diffractive amplitude on the free nucleon is obtained from
\begin{eqnarray}
\mathcal{A}(\gamma^* N \rightarrow V N ;W,\bq) = i \int d^2\br \rho_{V \gamma}(\br,\bq)  \sigma(\br,\bq) \, .
\label{eq:amplitude_N}
\end{eqnarray}
In performing the nuclear averages, we will encounter the integrals (we follow closely \cite{Glauber:1970jm} in our notation)
\begin{eqnarray}
M(\bb_+,\bb_-) &=& \int d^2 \bc T_A(\bc) \Gamma_N(\bb_+ - \bc, \bb_- - \bc) \nonumber \\
\Omega(\bb'_+,\bb'_-;\bb_+,\bb_-) &=& \int d^2\bc T_A(\bc) \Gamma^*_N(\bb'_+ - \bc, \bb'_- - \bc) \Gamma_N(\bb_+ - \bc, \bb_- - \bc) 
\end{eqnarray}
The optical thickness $T_A(\bb)$ of a nucleus is
\begin{eqnarray}
T_A(\bb) = \int_{-\infty}^{\infty} dz n_A(z,\bb) \, , \, \int d^2\bb \, T_A(\bb) = A \, ,
\end{eqnarray}
where the nuclear matter density $n_A(z,\bb)$ can be obtained from the standard Woods-Saxon parametrization for
heavy nuclei.
Now the function $\mathcal{C}$ obtains the form
\begin{eqnarray}
\mathcal{C}(\bb'_+,\bb'_-;\bb_+,\bb_-) &=& \Big[ 1 - {1 \over A} \Big(M^*(\bb'_+,\bb'_-) + M(\bb_+,\bb_-) \Big) + {1 \over A} \Omega(\bb'_+,\bb'_-;\bb_+,\bb_-) \Big]^A
\nonumber \\
&& - \Big[ \Big(1 -{1 \over A} M^*(\bb'_+,\bb'_-) \Big) \Big( 1- {1 \over A} M(\bb_+,\bb_-) \Big) \Big]^A
\label{eq:function_C}
\end{eqnarray}
Even at this level, of fully uncorrelated nucleons, the evaluation of the impact parameter integrals in eqs. \ref{eq;dsig_dt},\ref{eq:Sigma} 
is extremely cumbersome, and we will follow Glauber and Matthiae and study separately the limits of large $\bDelta^2 \gg R_A^{-2}$ and 
small $\bDelta^2 \lsim R_A^{-2}$ momentum transfers.
Let us start by expanding in the difference between the two terms in \ref{eq:function_C}, which then becomes
\begin{eqnarray}
\mathcal{C}(\bb'_+,\bb'_-;\bb_+,\bb_-) &=& \Big[ 1 - {1 \over A} \Big(M^*(\bb'_+,\bb'_-) + M(\bb_+,\bb_-) \Big) \Big]^{A-1}
\nonumber \\
&&\times \Big\{     \Omega(\bb'_+,\bb'_-;\bb_+,\bb_-) - {1 \over A}   M^*(\bb'_+,\bb'_-)M(\bb_+,\bb_-)  \Big\}    
\label{eq:function_C_2}
\end{eqnarray}
Now we observe, that in the limit $B_D \ll R_A^2$ (the nucleon profile in impact parameter space is much steeper than the one of the nucleus), 
so that we can simplify
\begin{eqnarray}
M(\bb_+,\bb_-) &\approx& {1 \over 2} \sigma(\br,0) T_A(\bb) , \nonumber \\
\Omega(\bb'_+,\bb'_-;\bb_+,\bb_-) &\approx& T_A({\bb+\bb' \over 2}) {1 \over 16 \pi^2} \, \int d^2\bq \exp[-i\bq(\bb'-\bb)] \sigma^*(\br',\bq) \sigma(\br,\bq) 
\nonumber \\
&\equiv& T_A({\bb+\bb' \over 2}) \chi(\br,\br',\bb-\bb')
\end{eqnarray}
In eq.\ref{eq:function_C_2}, the first factor then accounts for intranuclear absorption of color dipoles in the amplitude and its complex conjugate,
while the second factor is a single scattering  with coherent subtraction. The function $\Omega$ describes a single scattering off
a nucleon in the amplitude, and off the same nucleon in the conjugate amplitude.
We now freely use exponentiation, i.e. $(1 +x/A)^A \sim \exp(x)$ for $A \gg 1$, which is known to work well for medium to
heavy nuclei. Then, we come to the result for the incoherent cross section as
\begin{eqnarray}
{d\sigma_{\rm incoh} \over d\bDelta^2} &=& {1 \over 16 \pi} \Big\{ \int d^2\bb T_A(\bb) \Big| \int d^2\br \rho(\br,\bDelta) \sigma(\br,\bDelta) 
\exp[- \half \sigma(\br) T_A(\bb)] \Big|^2  \nonumber \\
&&- {1 \over A} \Big| \int d^2\bb \exp[-i \bb \bDelta] T_A(\bb) \int d^2\br \rho(\br,\bDelta) \sigma(\br,0) \exp[- \half \sigma(\br,0) T_A(\bb)]
\Big|^2 \Big\}.
\nonumber \\
\label{eq:dsig_lowt}
\end{eqnarray}
If we were to neglect intranuclear absorption, we would obtain for small $\bDelta^2$:
\begin{eqnarray}
{d\sigma_{\rm incoh} \over d\bDelta^2} = A \cdot {d\sigma (\gamma N \to V N) \over d\bDelta^2}\Big|_{\bDelta^2 = 0} 
\cdot \Big\{ 1- \mathcal{F}_A(\bDelta^2) \Big\} \, . 
\end{eqnarray}
As the nuclear formfactor 
\begin{eqnarray}
 \mathcal{F}_A(\bDelta^2)  = { 1 \over A} \int d^2\bb \, \exp[-i \bb\bDelta] T_A(\bb) 
\end{eqnarray} 
is unity at zero momentum transfer, the incoherent cross section in this (weak-scattering) approximation will vanish
in the forward direction. 
Evidently, the absorption factors in \ref{eq:dsig_lowt} will spoil this cancellation, and the incoherent cross section will have
a forward dip, but not vanish at $\bDelta^2 = 0$. 


In the limit of large momentum transfer, $\bDelta^2 \gg  R_A^{-2}$, the coherent subtraction will be suppressed 
by the quickly oscillating exponential. 
On the other hand it should be more appropriate
to sum up multiple quasielastic scatterings of the dipole. This amounts to taking all powers 
of $\Omega$ in eq. \ref{eq:function_C} into account and leads to
\begin{eqnarray}
\Sigma_{\rm incoh} (\br,\br',\bDelta) &=& \int d^2\bb d^2\bs \exp[ -i\bDelta \bs] 
\exp\Big[-\half \Big(\sigma^*(\br',0) + \sigma(\br,0) \Big)T_A(\bb) \Big] 
\nonumber \\
&&\times \Big\{ \exp[T_A(\bb) \chi(\br,\br',\bs)] -1 \Big \} \nonumber \\ 
&=& \sum_n {1 \over n!} \int d^2\bb T_A^n(\bb)  \exp\Big[-\half \Big(\sigma^*(\br',0) + \sigma(\br,0) \Big)T_A(\bb) \Big] \nonumber \\
&&
\int d^2 \bs \exp[ -i\bDelta \bs]  \chi^n(\br,\br',\bs)  \, .
\label{eq:Sigma_MS}
\end{eqnarray}
This result is the counterpart of eq. 19  in \cite{Kolya_DSE} and gives rise to the multiple scattering 
expansion of the coherent cross section:
\begin{eqnarray}
{d\sigma_{\rm incoh} \over d\bDelta^2} = {1 \over 16 \pi} \sum_n { 1 \over n!} \int d^2\bb T^n_A(\bb) \int \Big[ \prod_{i=1}^n d^2 \bq_i \Big] 
\delta^{(2)}(\bDelta - \sum_{i=1}^n \bq_i)  \Big|\mathcal{A}^{(n)}(\bDelta; \bq_1, \dots, \bq_n) \Big|^2 \, .
\nonumber \\
\label{eq:MSE}
\end{eqnarray}
Here we introduced an effective amplitude of $n$ quasielastic rescatterings in the presence of nuclear absorption
\begin{eqnarray}
{\mathcal A}^{(n)}(\bDelta,\bq_1,\dots,\bq_n) = { 1 \over (4 \pi)^{n-1}}  \, \int d^2\br \rho(\br,\bDelta) 
\exp[-\half \sigma(\br) T_A(\bb)] 
\sigma(\br,\bq_1) \dots \sigma(\br,\bq_n)
\label{eq:MSE_amp}
\end{eqnarray}
A few comments on the coherence pattern of the multiple scattering expansion \ref{eq:MSE} are in order, see
also the corresponding discussion in \cite{Kolya_DSE}. 

Firstly, we work in a regime where the lifetime of the 
$Q \bar Q$ dipole and the formation time of the final state vector mesons are much larger than the radius $R_A$
of the nucleus. Correspondingly, the factor $\exp[-\sigma(\br)T_A(\bb)/2]$ in \ref{eq:MSE_amp} describe the attenuation 
of the dipole wave {\it{amplitude}} before and after the quasielastic scatterings. 
Notice that they are responsible for the survival of the large rapidity gap between the vector meson and the 
reaction products in the nuclear fragmentation region. They affect the ``hard scattering'' quasielastic  {\it{amplitude}} in a nontrivial way.
A universal ``gap survival probability'' which simply multiplies a Born-level cross section is not borne out by the quantum
level treatment of the production process. 

Secondly, in the $n$-th term of the multiple scattering expansion, the color dipole receives $n$ kicks
$\bq_1, \dots, \bq_n$ which add up to the total momentum transfer $\bDelta$. There is no interference
between different numbers of scatterings. We recall the well known analogy of the quasielastic scattering on
nuclei \cite{Glauber:1970jm} with the parton model of deep inelastic scattering: the factor $\propto T_A^n(\bb)$
signifies an n-nucleon density in the nucleus, i.e. the {\it probability} to find $n$ nucleons participating in
the quasielastic scattering.

Thirdly, also scatterings at different impact parameters only interfere, if they
take place within a tube of cross section $\sim B$, the effective area of the dipole-nucleon interaction. 
To the extent that the latter can be neglected against the nuclear size (which we assume), different impact
parameters add up incoherently. Higher order terms in the expansion \ref{eq:MSE} increase in importance at larger momentum
transfers $\bDelta$, as the multiple convolution over the $\bq_n$ get broader and broader with increasing order.
As the size of the interaction region in the elementary diffractive process increases with energy 
(``shrinkage of the diffraction cone'') one may envision a limit where it becomes similar to, or exceeds,
the nuclear size. In such a limit the quasielastic momentum transfers may become so small that they would
not break a nucleus \cite{Zoller:1988gz}. We do not discuss such a possibility in this work. 

\section{Numerical results}

\subsection{$\gamma A$-collisions}

We start from the $\gamma A \to V X$ reaction, which may be studied at a future electron-ion collider, see
e.g. \cite{Caldwell:2010zza,Lappi:2014foa} for interesting suggestions. 
We first wish to analyze the importance of multiple scattering contributions in the expansion 
\ref{eq:MSE}, \ref{eq:MSE_amp}. 
To make the problem numerically more tractable, we start by assuming for the off-forward dipole cross section 
a factorized ansatz (we from now on put the dependence on $x = m_V^/W^2$ in evidence, where $m_V$ is the mass
of the vector meson):
\begin{eqnarray}
\sigma(x,\br, \bq) = \sigma(x,r) \exp[-\half B \bq^2] \,  ,
\end{eqnarray}
valid within the forward diffractive cone,
with a diffractive slope $B$, which depends on the vector meson and on energy.
We the readily obtain for the multiple scattering expansion of the
incoherent cross section
\begin{eqnarray}
{d \sigma_{\rm incoh}\over d\bDelta^2} = \sum_n {d \sigma^{(n)} \over d\bDelta^2} =
{1 \over 16 \pi} \sum_n  w_n(\bDelta) \int d^2\bb T^n_A(\bb) |I_n(x,\bb)|^2 \, ,
\label{eq:MSE_approx}
\end{eqnarray}
where the transverse momentum dependent coefficients are
\begin{eqnarray}
w_n(\bDelta) = {1 \over n \cdot n!} \cdot \Big({1 \over 16 \pi B}\Big)^{n-1} \cdot \exp\Big(-{B\over n} \bDelta^2 \Big) \, ,
\end{eqnarray}
and
\begin{eqnarray}
I_n(x,\bb) &=& \bra{V}  \, \sigma^n(x,r) \exp[-\half \sigma(x,r) T_A(\bb)] \ket{\gamma} \nonumber \\
&=& \int_0^1 dz \int d^2\br \, \Psi^*_V(z,\br) \Psi_\gamma(z,\br) \, \sigma^n(x,r) \exp[-\half \sigma(x,r) T_A(\bb)] \, . 
\label{eq:In}
\end{eqnarray}
We neglect the $\bDelta$-dependent phase factor in eq. \ref{eq:overlap}, and use the vector meson wave function 
derived from the $\gamma_\mu$ vertex. Both these approximations are valid in a nonrelativistic limit and are admissible,
as long as one is not interested in subtle relativistic effects such as helicity flip amplitudes. 
The overlap of vector meson and photon light-cone wave function is then \cite{Nemchik:1994fp,Nemchik:1996cw}
\begin{eqnarray}
\Psi^*_V(z,\br) \Psi_\gamma(z,\br) &=& { e_Q \sqrt{4 \pi \alpha_{\rm em} } N_c \over 4 \pi^2 z(1-z)} \Big\{ m_Q^2 K_0(m_Q r) \psi(z,r) 
\nonumber \\
&&- [z^2 + (1-z)^2] m_QK_1(m_Q r) 
	{\partial \psi(z,r) \over \partial r}   \Big\} .
\end{eqnarray}
For the radial wave function $\psi(z,r)$, we choose the so-called ``boosted Gaussian'' wave function 
\cite{Nemchik:1994fp,Nemchik:1996cw} as parametrized in 
\cite{Kowalski:2006hc} and \cite{Cox:2009ag} for the $J/\psi$ and $\Upsilon$ meson, respectively. 

\begin{figure}[!h]
	\begin{center}
		\includegraphics[width=.4\textwidth]{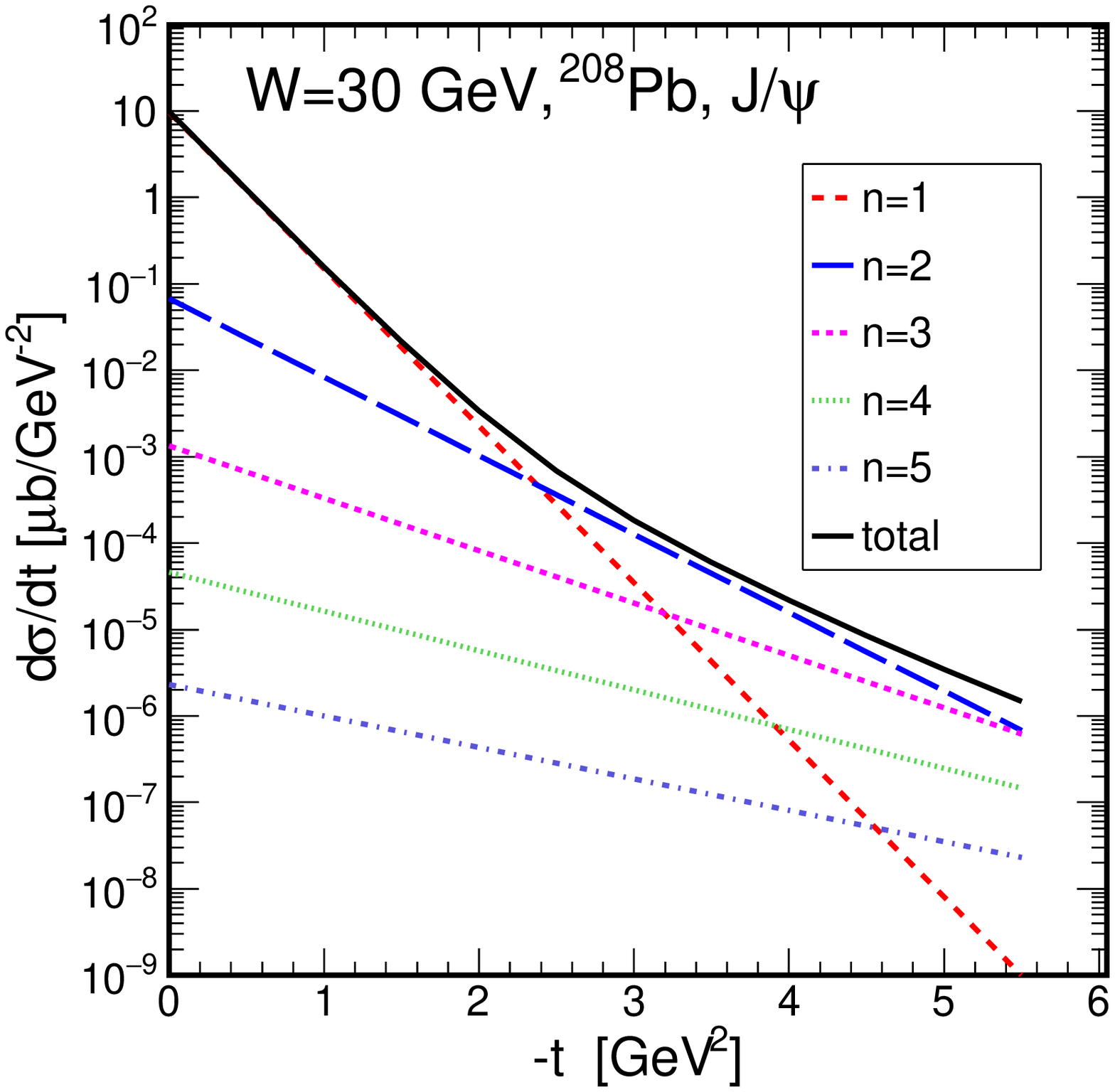}
		\includegraphics[width=.4\textwidth]{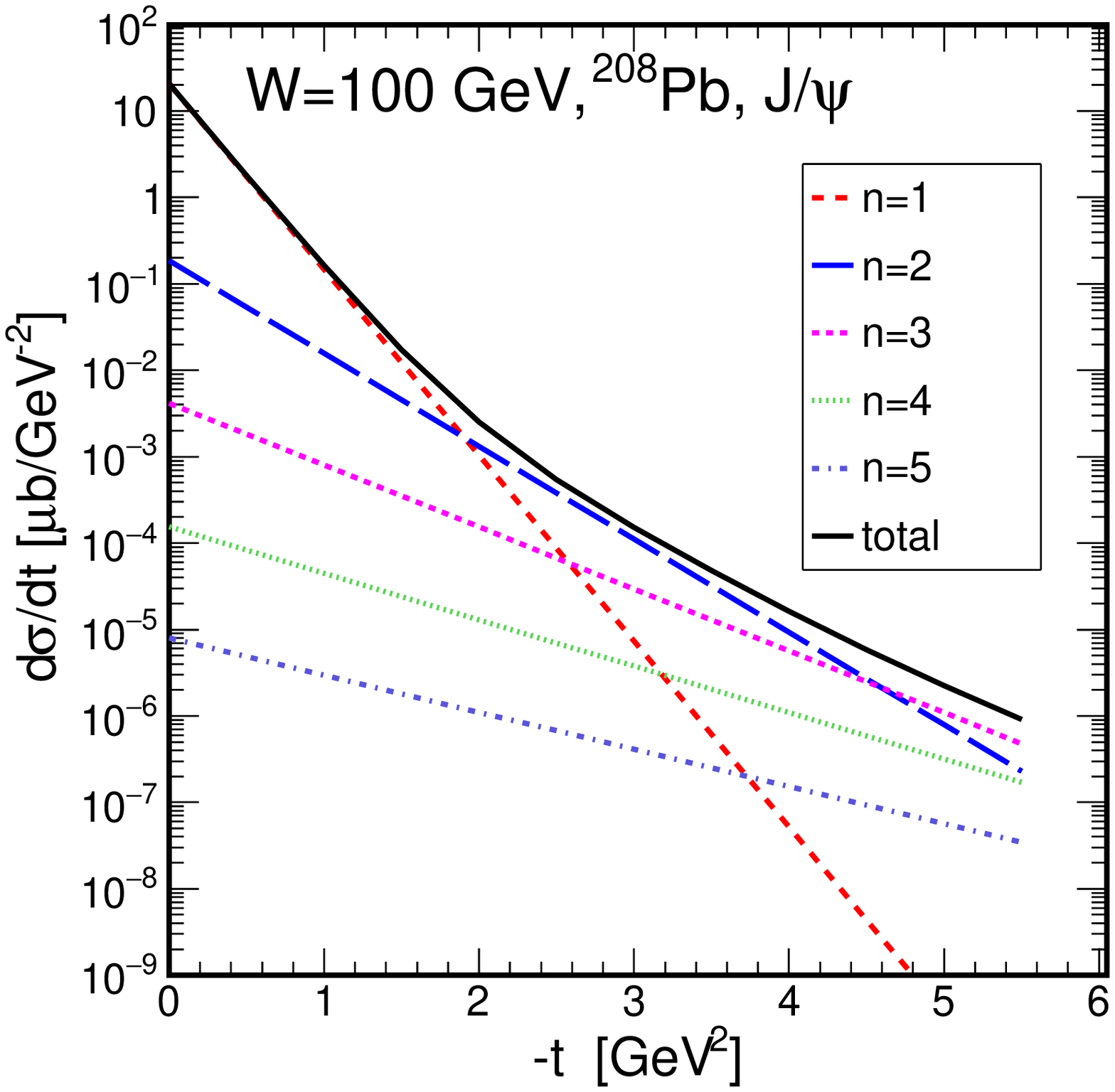}
		\caption{Incoherent diffractive cross section for $\gamma A \to J/\psi X$ for
			${A=^{208}{\rm Pb}}$ at $W=30 \, \rm{GeV}$ (left panel)  and $W=100 \, \rm{GeV}$ (right panel). 
			Shown are the contributions from 1 to 5 scatterings.
		}
		\label{fig:dsig_dt_Pb_Jpsi}
	\end{center}
\end{figure}

\begin{figure}[!h]
	\begin{center}
		\includegraphics[width=.4\textwidth]{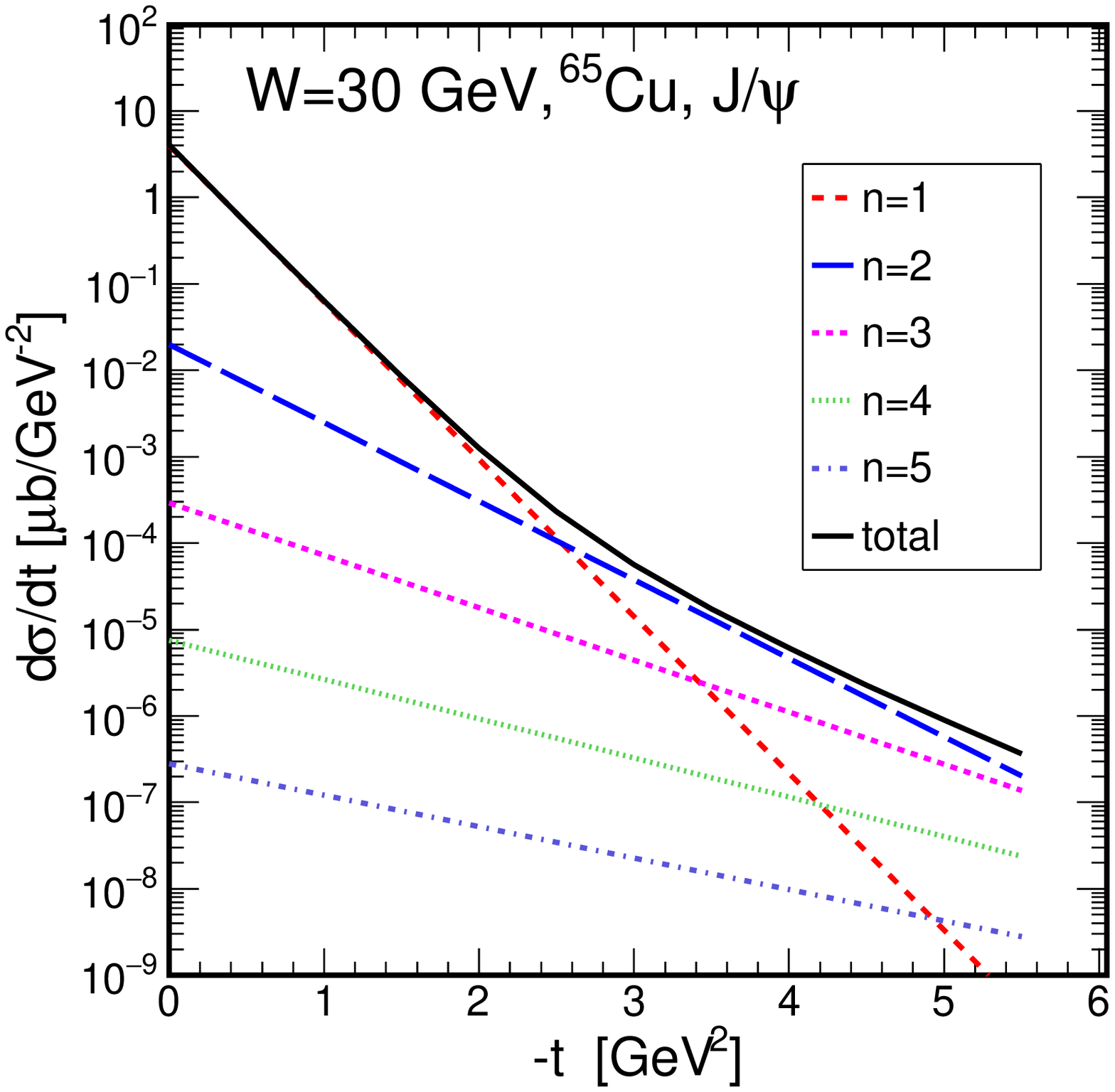}
		\includegraphics[width=.4\textwidth]{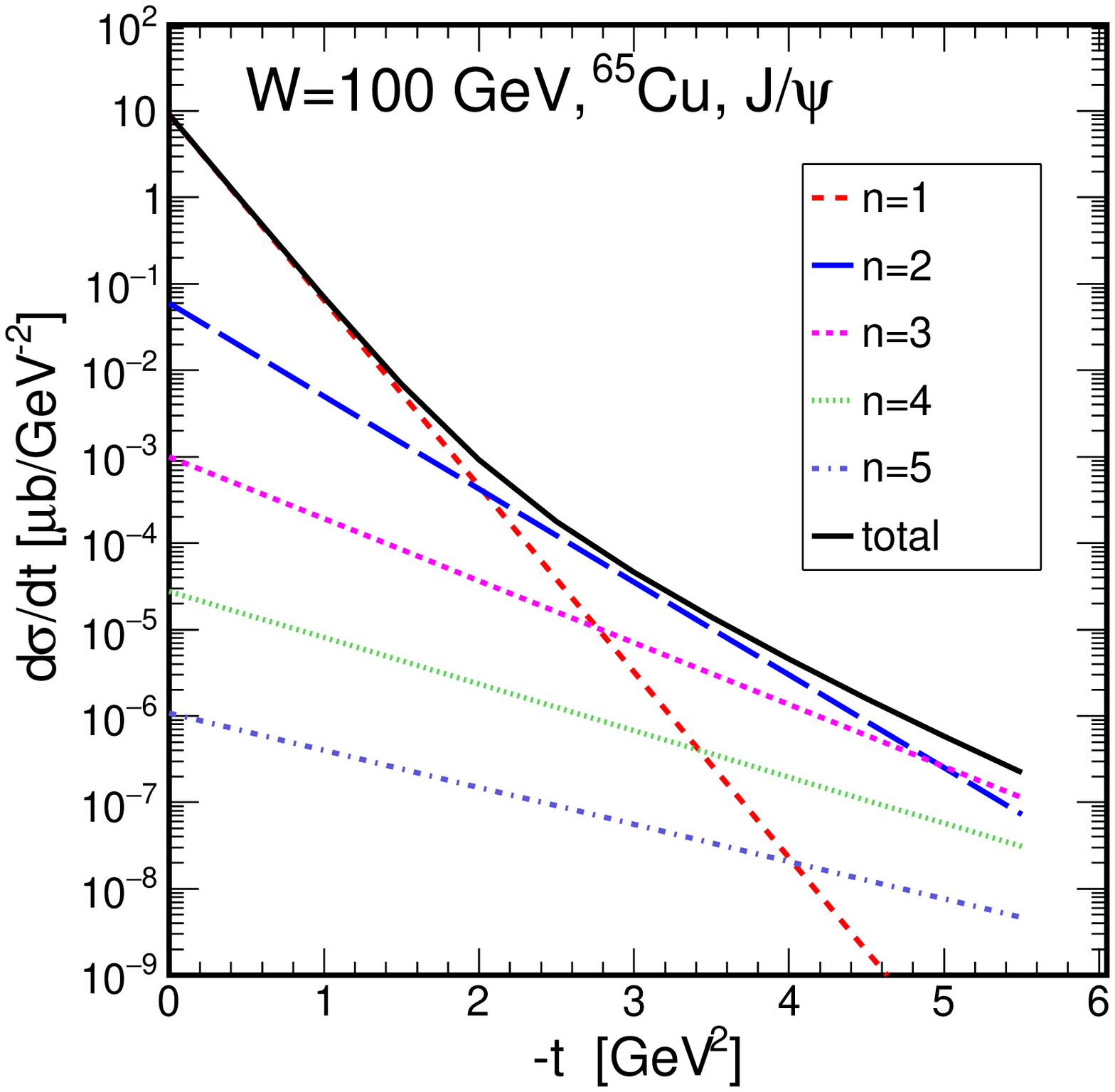}
		\caption{Incoherent diffractive cross section for $\gamma A \to J/\psi X$ for
			${A=^{65}{\rm Cu}}$ at $W=30 \, \rm{GeV}$ (left panel)  and $W=100 \, \rm{GeV}$. 
			Shown are the contributions from 1 to 5 scatterings.
		}
		\label{fig:dsig_dt_Cu_Jpsi}
	\end{center}
\end{figure}

\begin{figure}[!h]
	\begin{center}
		\includegraphics[width=.4\textwidth]{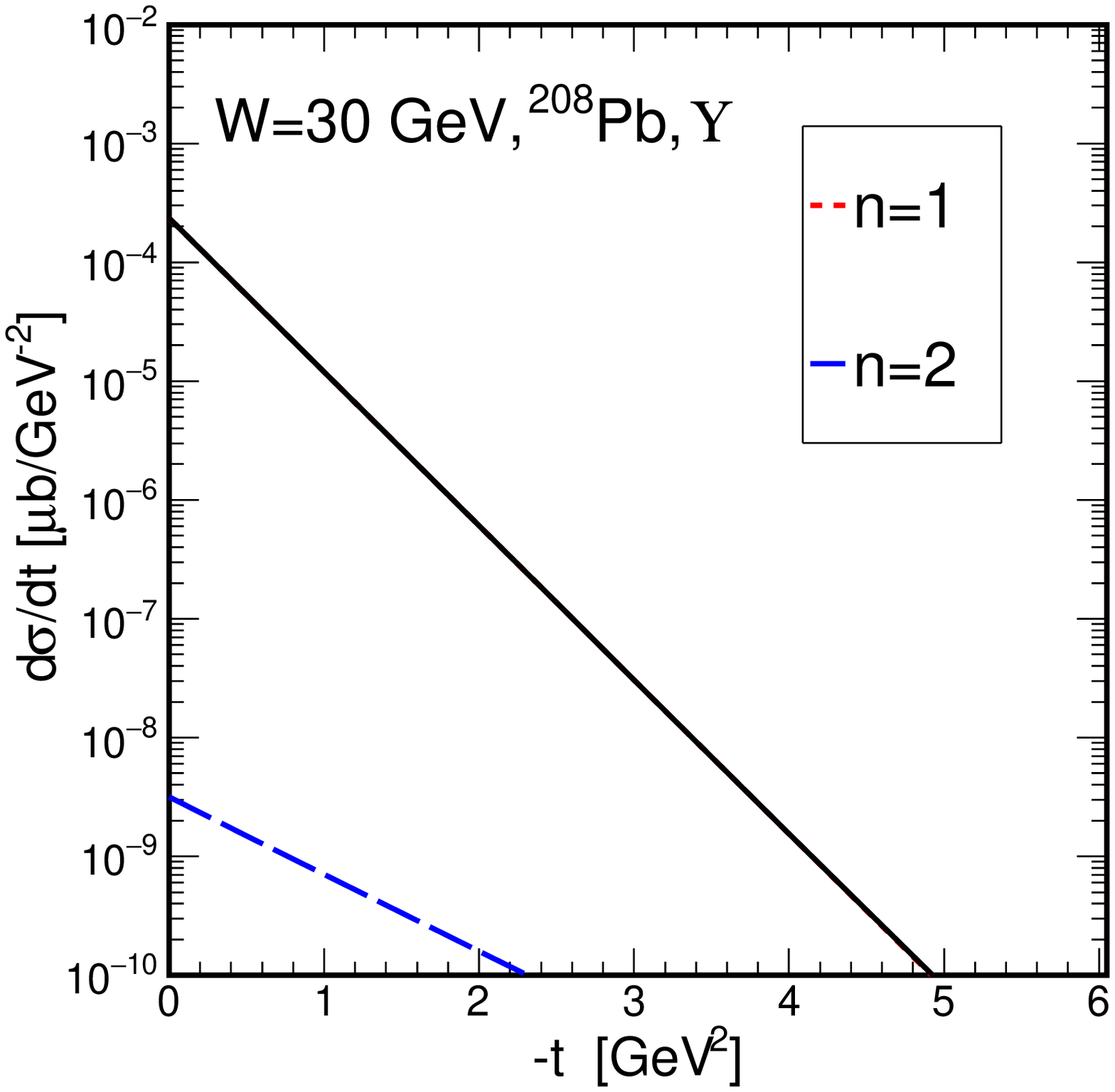}
		\includegraphics[width=.4\textwidth]{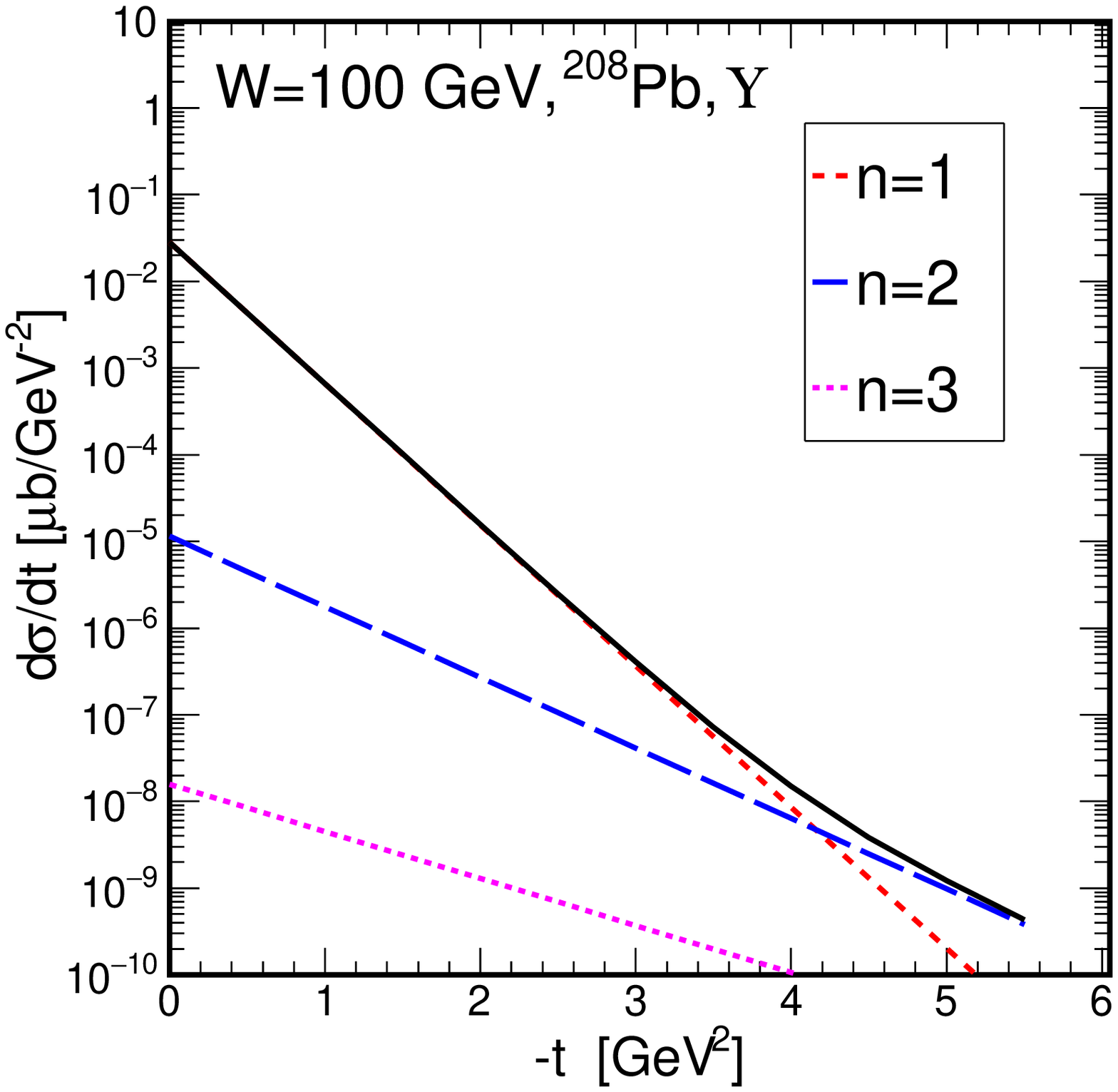}
		\caption{Incoherent diffractive cross section for $\gamma A \to \Upsilon X$ for
			${A=^{208}{\rm Pb}}$ at $W=30 \, \rm{GeV}$ (left panel)  and $W=100 \, \rm{GeV}$. 
			Shown are the contributions from 1 to 2 (left panel), respectively 3 (right panel)  scatterings.
		}
		\label{fig:dsig_dt_Pb_Ups}
	\end{center}
\end{figure}

\begin{figure}[!h]
	\begin{center}
		\includegraphics[width=.4\textwidth]{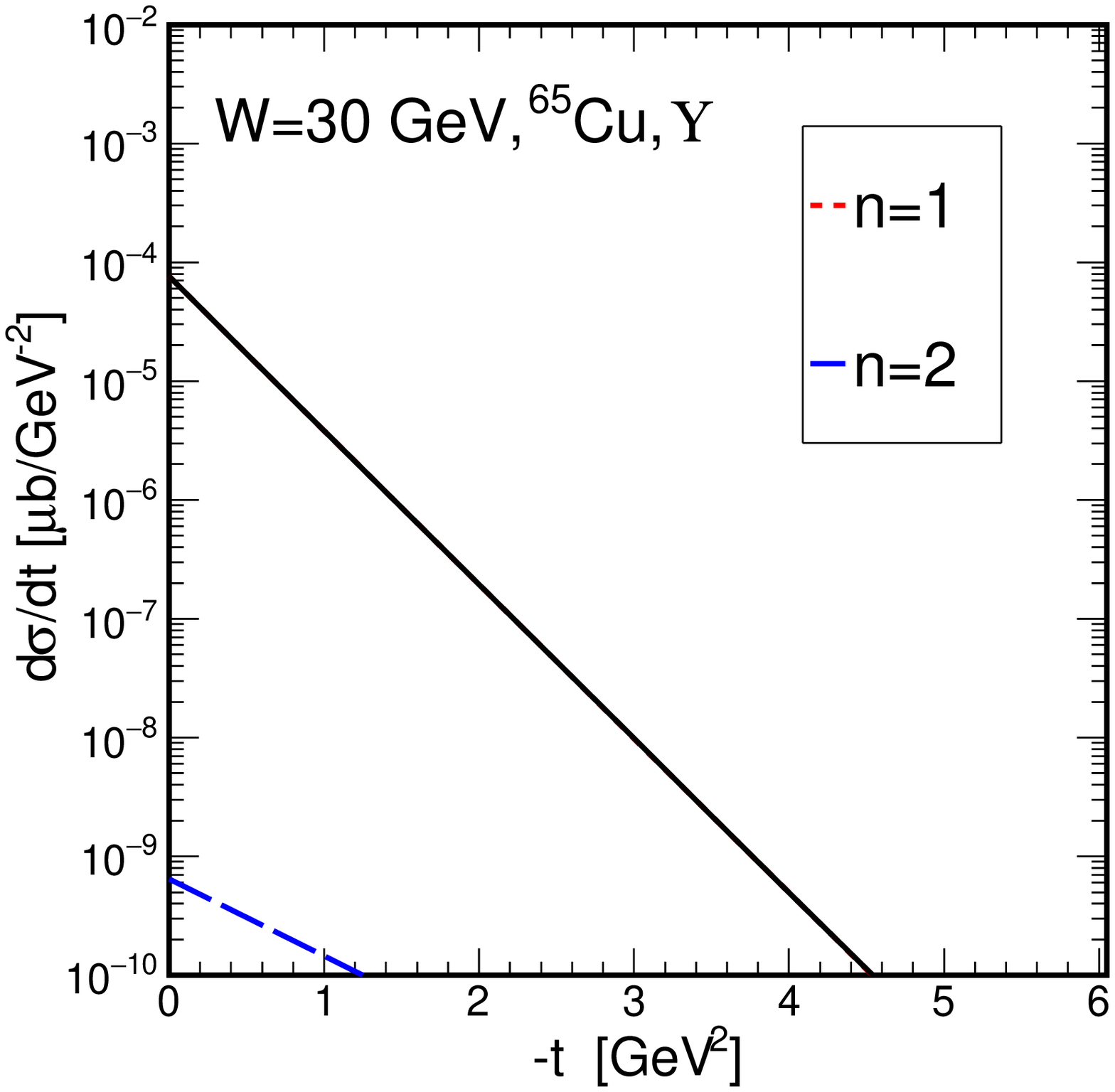}
		\includegraphics[width=.4\textwidth]{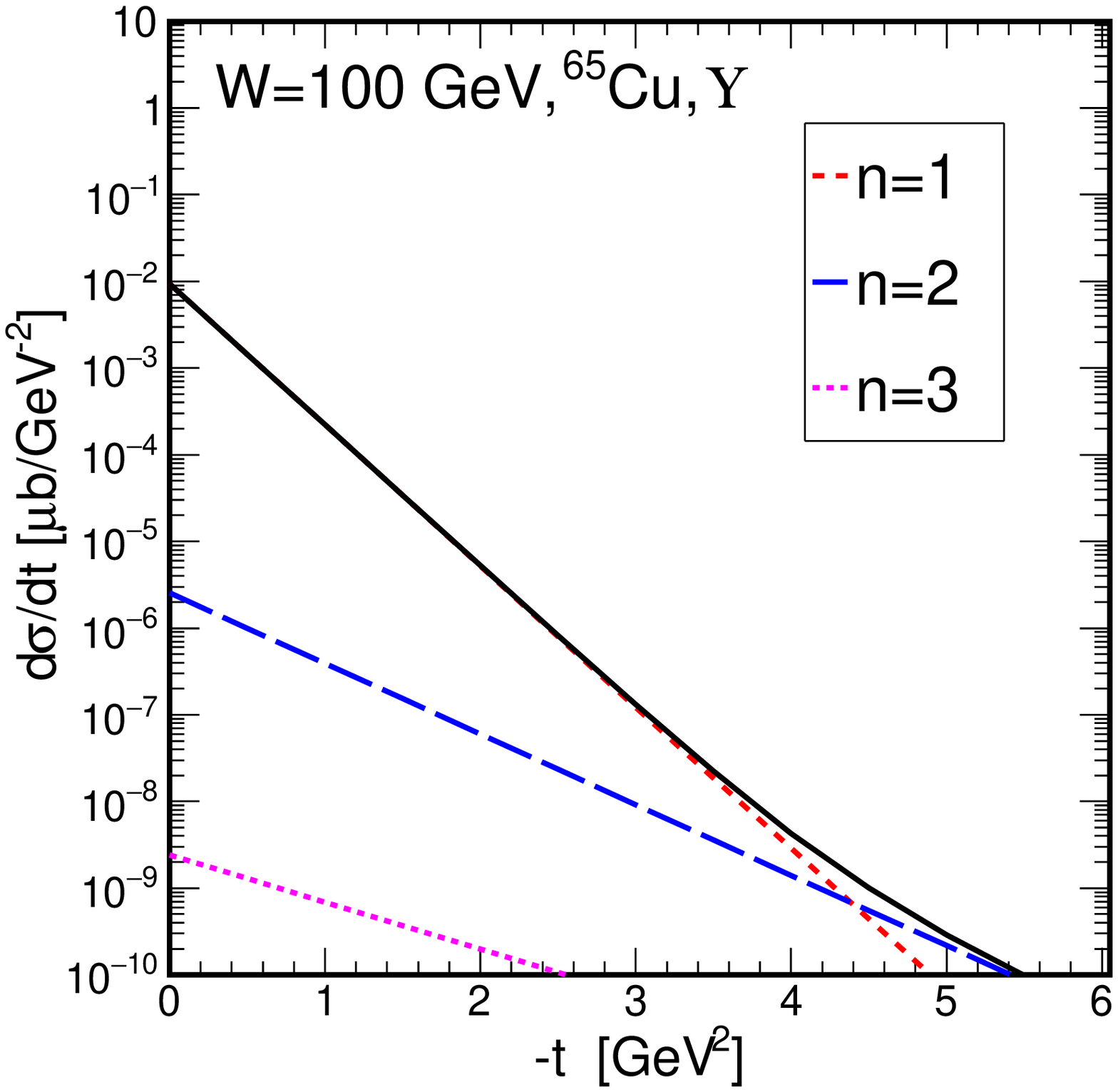}
		\caption{Incoherent diffractive cross section for $\gamma A \to \Upsilon X$ for
			${A=^{65}{\rm Cu}}$ at $W=30 \, \rm{GeV}$ (upper panel)  and $W=100 \, \rm{GeV}$. 
			Shown are the contributions from 1 to 2 (left panel), respectively 3 (right panel) scatterings.
		}
		\label{fig:dsig_dt_Cu_Ups}
	\end{center}
\end{figure}

Some comments on the dipole cross section are in order. We will use color dipole cross sections fitted to
the proton structure functions measured at HERA using the representation of total photoabsorption
cross sections \cite{Nikolaev:1990ja}
\begin{eqnarray}
\sigma_{T,L}(\gamma^*(Q^2) p) = \int_0^1 dz \int d^2\br \Big| \Psi^{\gamma^*}_{T,L}(z,\br,Q^2) \Big|^2 \, \sigma(x,r) \, .
\end{eqnarray}

The ansatz of the fit for $\sigma(x,r)$ follows an improvement of the Golec-Biernat and W\"usthoff (GBW) model \cite{GolecBiernat:1998js} 
proposed by Bartels, Golec-Biernat and Kowalski, (BGK)  ~\cite{Bartels:2002cj}, which takes into account the  DGLAP evolution of the gluon density 
in an explicit way. The model preserves the GBW eikonal approximation to saturation and thus the dipole cross section is given by
\begin{equation}
\label{eBGK}
\sigma(x,r) = \sigma_{0} \left(1 - \exp \left[-\frac{\pi^{2} r^{2} \alpha_{s}(\mu^{2}) xg(x,\mu^{2})}{3 \sigma_{0}} \right]\right).
\end{equation}
The evolution scale $\mu^{2}$ is connected to the size of the dipole by $\mu^{2} = C/r^{2}+\mu^{2}_{0}$. 
This assumption allows to treat  consistently the contributions of large dipoles without making 
the strong coupling constant, $\alpha_s(\mu^2$), un-physically large. 
This means also that we can extend the model, keeping its perturbative character, to the data at low $Q^2$, 
because the external $Q^2$ and the internal $\mu^2$ scales are connected only by the wave function.  
The gluon density, which  is parametrized  at the starting scale $\mu_{0}^{2}$, 
is evolved to larger scales, $\mu^2$, using NLO DGLAP evolution.
At the starting scale $\mu_0^2 = 1.9 \, \rm{GeV}^2$ two different forms of the gluon density 
were explored \cite{Luszczak:2013rxa,Luszczak:2016bxd}:
\begin{itemize}
	\item 
	the {\it soft} ansatz, as used in the original BGK model 
	\begin{equation}
	xg(x,\mu^{2}_{0}) = A_{g} x^{-\lambda_{g}}(1-x)^{C_{g}},
	\label{gden-soft}
	\end{equation}
	\item
	the {\it soft + hard} ansatz
	\begin{equation}
	xg(x,\mu^{2}_{0}) = A_{g} x^{-\lambda_{g}}(1-x)^{C_{g}}(1+D_g x +E_gx^2),
	\label{gden-softhard}
	\end{equation}
\end{itemize}
The fit was performed using the xFitter framework \cite{xfitter}, which contains the newest HERA data.
For details, see \cite{Luszczak:2013rxa,Luszczak:2016bxd}.
Below, we will use the following fits:
\begin{itemize}
	\item fit I:
BGK fit with fitted valence quarks for $\sigma_r$  for H1ZEUS-NC data in the range $Q^2 \ge 3.5$~GeV$^2$  and $x\le 0.01$. NLO fit. 
{ \it Soft gluon}.   $ m_{uds}= 0.14, m_{c}=1.3$ GeV.  $Q_0^2=1.9$ GeV$^2$. It is the fit from Table 2 (fit I) \cite{Luszczak:2016bxd} 

\item fit II: BGK fit with valence quarks for $\sigma_r$  for H1ZEUS-NC data in the range $Q^2 \ge 0.35$~GeV$^2$ and $x\le 0.01$. NLO fit.  
{ \it Soft + hard gluon}.   $ m_{uds}= 0.14, m_{c}=1.3$ GeV, saturation ansatz. $Q_0^2=1.9$. It is the fit from Table 11 (fit II) from \cite{Luszczak:2016bxd}.
\end{itemize}

Let us turn to the results for $t$-distributions. In these calculations we used dipole fit II.
In fig.\ref{fig:dsig_dt_Pb_Jpsi} we show the incoherent diffractive photoproduction cross section for $J/\psi$-mesons
for two different energies, $W=30 \, \rm{GeV}$ and $W=100 \, \rm{GeV}$, corresponding to $x = 0.01$ and $x=0.001$ respectively.
on a $^{208}\rm{Pb}$ nuclear target. We show the contribution of up to five scatterings. 
We see that the differential cross section deviates from the exponential shape as is expected from 
the superposition of exponentials with different slope.
For $W=30 \, {\rm GeV}$ the double scattering takes over only at a large value of $|t| \gsim 2.5 \, \rm{GeV}^2$, 
beyond the diffraction cone of the free nucleon process, and the triple scattering only at very large $|t| \gsim 4.8 \, \rm{GeV}^2$.
The crossing of single and double scattering moves to a slightly lower $|t| \sim 2.2 \, \rm{GeV}^{2}$ when the energy inreases to
$W= 100 \, \rm{GeV}$. 
To demonstrate the evolution with nuclear size of the observable, we show in fig. \ref{fig:dsig_dt_Cu_Jpsi} the $t$-distribution
for the $^{65}{\rm Cu}$ target, for the same two energies. The qualitative picture is very similar to the heavier lead nucleus.

The diffractive production of $\Upsilon$ involves much smaller dipole sizes, and one would expect weaker nuclear effects
in this case. As an example of the weak nuclear attenuation limit, we show distributions for $\Upsilon$ production in fig.
\ref{fig:dsig_dt_Pb_Ups} for the lead target and in fig. \ref{fig:dsig_dt_Cu_Ups} for copper. Indeed here the effects of multiple 
scattering are delayed to very large values of $t$. At $W = 100 \, \rm{GeV}$ the double scattering crosses the single scattering
contribution only at $|t| \sim 4.5 \, \rm{GeV}^{2}$

\begin{figure}[!h]
	\begin{center}
		\includegraphics[width=.5\textwidth]{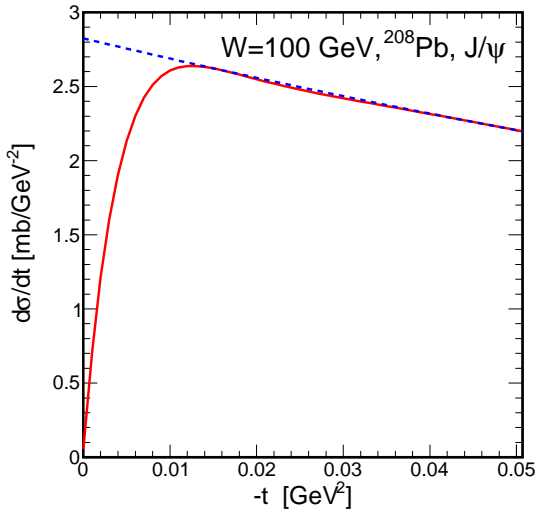}
		\caption{A zoom into the region of very low $t$ of the incoherent cross section. The solid line shows the forward dip, while
			the dashed line is the single scattering term of the large-$t$ limit.}
		\label{fig:low_t}
	\end{center}
\end{figure}

We see, that the single scattering term generally dominates for heavy vector mesons over a broad range
of $t$. For very small $\bDelta^2$, as discussed in section \ref{sec:formalism}, we should rather take 
the absorption improved single-scattering term calculated from
\begin{eqnarray}
{d \sigma_{\rm incoh}\over d\bDelta^2} = 
{1 \over 16 \pi} \Big\{ w_1(\bDelta) \int d^2\bb T_A(\bb) |I_1(x,\bb)|^2 - {1 \over A} \Big| \int d^2\bb \exp[-i \bDelta \bb] T_A(\bb) I_1(x,\bb) \Big|^2 
\Big\}\, . \nonumber \\
\label{eq:low_t}
\end{eqnarray}
As an example, we show the result for the resulting $t$-dependence in fig \ref{fig:low_t}  for $J/\psi$ production at $W=100 \, \mathrm{GeV}$ on a 
$^{208}Pb$ target zooming into the region of very small $|t|$.
Also shown is the first term of eq. \ref{eq:low_t}, which coincides with the single scattering term of eq. \ref{eq:MSE_approx}. 
We see, that the result of eq. \ref{eq:low_t} has a sharp dip in forward direction, converges rather quickly to the large-$t$ single scattering 
cross section.  Within the approximations adopted in this section it is therefore not necessary to discuss separately
a region of intermediate $t$. 
Due to the sharpness of the forward dip, the single scattering approximation of eq \ref{eq:MSE_approx} is an appropriate approximation
for the $t$-values to be of interest e.g. in ultraperipheral heavy-ion collisions.

\begin{figure}[!h]
 	\begin{center}
 		\includegraphics[width=.5\textwidth]{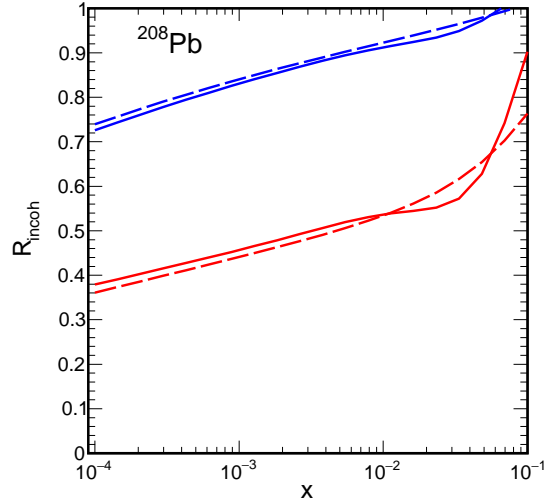}
 		\caption{The nuclear suppression ratio (\ref{eq:ratio}) for $\gamma A \to V X$ on lead as 
 			a function of $x = m_V^2 /W^2$ for $V = \Upsilon$ (upper lines) and $V=J/\psi$ (lower lines).
 		The dashed lines refer to dipole cross section fit I, while the solid lines show the result for fit II.}
 		\label{fig:Ratio_Pb}
 	\end{center}
 \end{figure}

In the absence of nuclear absorption, we would get the impulse approximation result
\begin{eqnarray}
{d \sigma^{\rm IA}_{\rm incoh}\over d\bDelta^2} = A \cdot {d \sigma (\gamma N \to V N) \over d\bDelta^2} = {w_1 (\bDelta) \over 16 \pi} \, 
\Big| \bra{V} \sigma(x,r) \ket{\gamma} \Big|^2  \, .
\end{eqnarray} 
A measure of the strength of nuclear absorption effects is the ratio \cite{Kopeliovich:1991pu}
\begin{eqnarray}
R_{\rm incoh}(x) = {{d \sigma_{\rm incoh} / d\bDelta^2} \over A \cdot {d \sigma (\gamma N \to V N) / d\bDelta^2} } =
 {\displaystyle \int d^2\bb T_A(\bb) \Big| \bra{V} \sigma(x,r) \exp[-\half \sigma(x,r) T_A(\bb)] \ket{\gamma}\Big|^2 \over \Big| \bra{V} \sigma(x,r) \ket{\gamma} \Big|^2} \, .
 \nonumber \\
 \label{eq:ratio}
\end{eqnarray}
We show this ratio as a function of $x$ in fig. \ref{fig:Ratio_Pb} for the $^{208}\rm{Pb}$ nucleus. Here the upper two lines shows the result
for $\Upsilon$ production and the lower two lines for $J/\psi$. 
The dashed lines show the result for dipole fit I, while the solid lines refer to dipole fit II. We see that the nuclear suppression
at small $x$ depends only weakly on the dipole cross section that was used. A substantial difference exists only at $x \gsim 0.01$, but
keep in mind that the dipole approach is by construction valid at small $x$. In practice $x \lsim 0.03$ is a conservative estimate of
its region of applicability.

Notice that although both $J/\psi$ and $\Upsilon$ production involve color dipoles small enough
to justify the use perturbative QCD, it is not admissible to neglect the nuclear attenuation in either case. 
Although the dipole cross section at the typical dipole size is rather small, the nuclear opacity is enhanced by the large nuclear
size.
 
\subsection{Ultraperipheral heavy-ion collisions}

Information on diffractive vector meson production on nuclear targets at the highest energies comes from
ultraperipheral heavy-ion collisions, see e.g. the reviews \cite{Contreras:2015dqa,Bertulani:2005ru,Baur:2001jj}.
The rapidity distribution of incoherently produced vector mesons can be calculated straightforwardly from
\begin{eqnarray}
{d\sigma_{\rm incoh} (AA \to V AX) \over dy} =  n_{\gamma/A}(z_+) \sigma_{\rm incoh}(W_+) + n_{\gamma/A}(z_-) \sigma_{\rm incoh}(W_-) \, ,
\end{eqnarray}
with 
\begin{eqnarray}
z_{\pm} = {m_V \over \sqrt{s_{NN}} } e^{\pm y },  \, \,  W_{\pm} = \sqrt{ z_{\pm} s_{NN}} \, . 
\end{eqnarray}
For our purposes it is sufficient to use the quasiclassical Weizs\"acker-Williams flux of photons with the 
``ultraperipheral'' requirement that the centers of colliding nuclei are at least a 
distance $2 R_A$ apart in impact parameter space. Explicit expressions are found e.g. in \cite{Baur:2001jj}.

\begin{figure}[!h]
	\begin{center}
		\includegraphics[width=.4\textwidth]{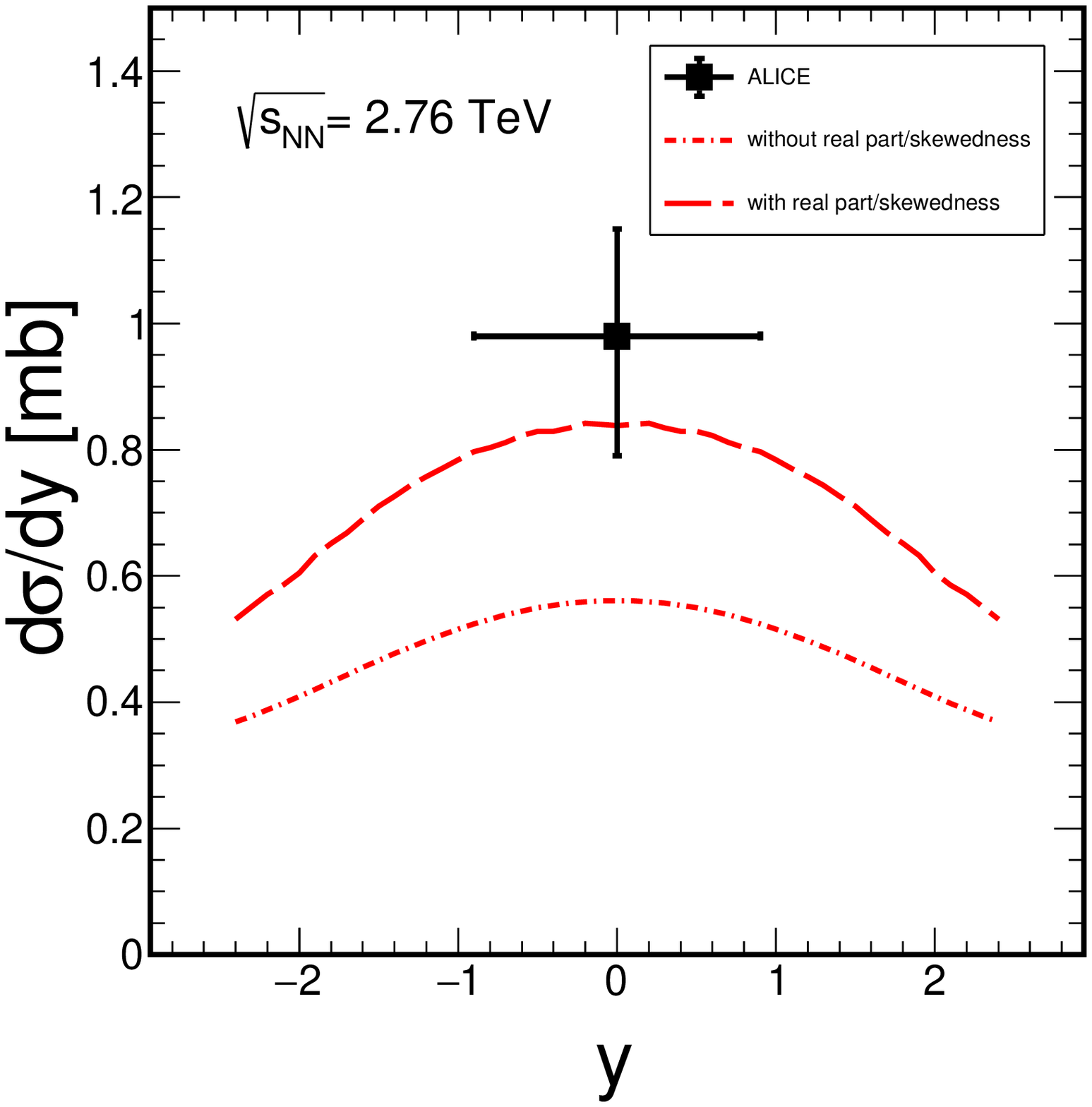}
		\includegraphics[width=.4\textwidth]{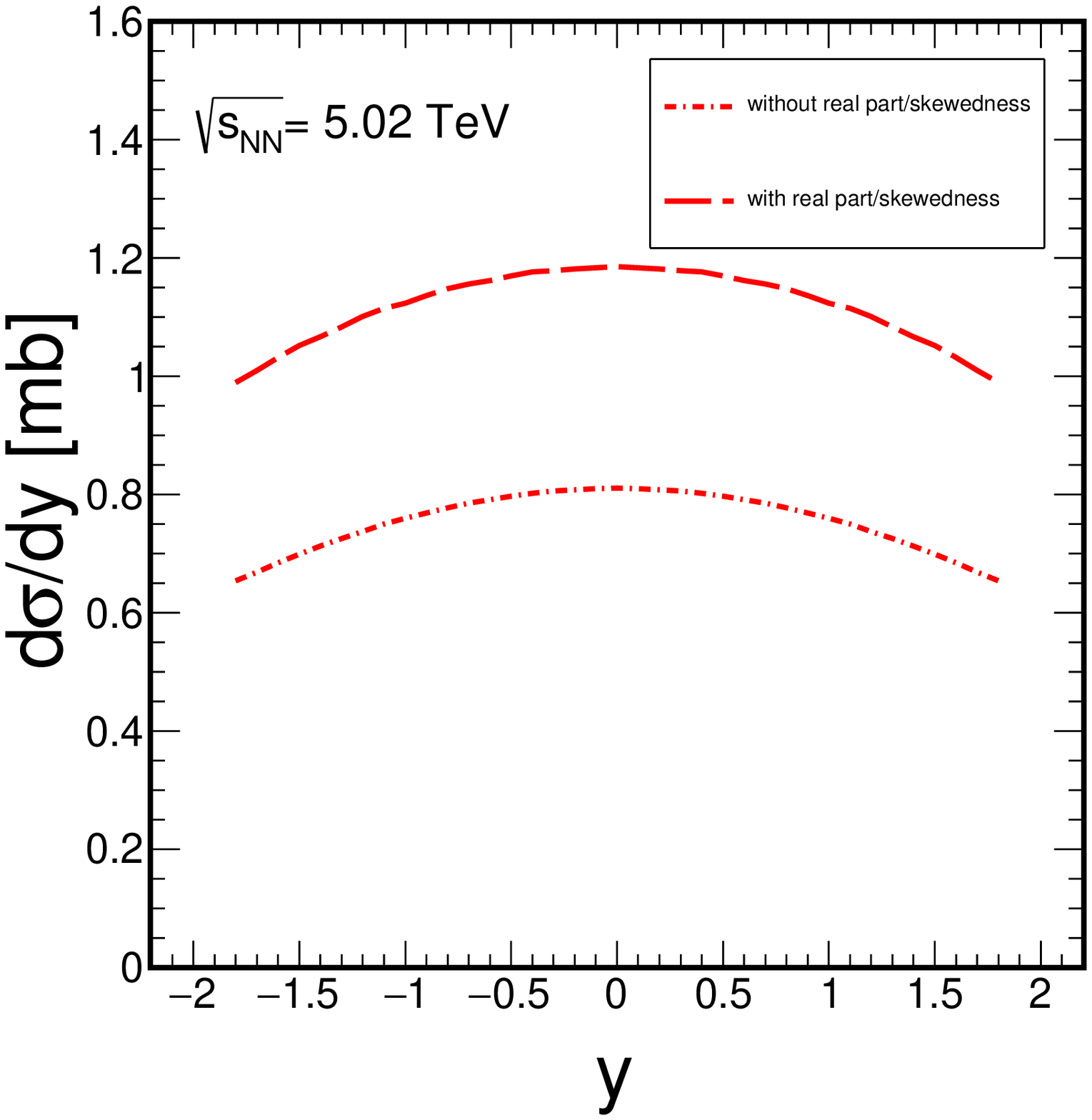}
		\caption{Incoherent diffractive cross section for $\gamma A \to J/\psi X$ for
			${A=^{65}{\rm Cu}}$ at $W=31 \, \rm{GeV}$ (upper panel) and $W=100 \, \rm{GeV}$. Shown are the contributions from 1 to 5 scatterings.
			Here the nuclear absorption was not taken into account.}
		\label{fig:dsig_dy_Jpsi}
	\end{center}
\end{figure}
Before comparing to experimental data, there are two omissions in the formalism presented above, 
which should be accounted for.

Firstly, we have throughout assumed, that the color dipole amplitude \ref{eq:dipole_amplitude} is purely absorptive.
Consequently, the photoproduction amplitude  \ref{eq:amplitude_N} is dominated by its imaginary part.
The structure of \ref{eq:function_C}, \ref{eq:function_C_2} strongly suggests, that only the absorptive part of the dipole
amplitude enters the nuclear suppression factors. It seems then reasonable, at least for the absorption-corrected 
nuclear amplitude to assume that the relevant part is the same as for the free nucleon amplitude.
Introducing the effective $x$-dependent intercept
\begin{eqnarray}
\Delta_\Pom = {\partial \log \Big( \bra{V} \sigma(x,r) \ket{\gamma} \Big) \over \partial \log(1/x)} \, , 
\end{eqnarray}
The real part is restored by replacing 
\begin{eqnarray}
\sigma(x,r) \to (1 - i \rho(x)) \sigma(x,r) \, , \, \rho(x) = \tan\Big( {\pi \Delta_\Pom \over 2} \Big) 
\end{eqnarray}

Secondly, the color dipole cross section has been obtained from a fit of the total photoabsorption cross
section on the nucleon, i.e. a fit to the absorptive part of the forward, $t=0$,  Compton amplitude. 
In vector meson production, even at $\bDelta =0$ the $t=0$ limit is not reached at finite energy 
there is always a finite $t_{\rm min}$ due to the vector meson mass.
The corresponding correction to the amplitude is Shuvaev's \cite{Shuvaev:1999ce} factor
\begin{eqnarray}
R_{\rm skewed} = { 2^{2 \Delta_\Pom + 3} \over \sqrt{\pi} } \cdot {\Gamma(\Delta_\Pom + 5/2) \over \Gamma(\Delta_\Pom+ 4)} \, .
\end{eqnarray}
This correction has been studied with some rigour only for the two-gluon ladder, where it accounts for the ``skewedness''
of gluon momentum fractions.

These two corrections were not important for the discussion of salient features of $t$-distributions nor for 
the nuclear suppression ratio discussed above. They do however affect the prediction for the cross section.
We apply them to our result by multiplying eq.\ref{eq:dsig_lowt} by
\begin{eqnarray}
K = (1 + \rho^2(x)) \cdot R^2_{\rm skewed} \, .
\label{eq:K-factor}
\end{eqnarray} 
In fig. \ref{fig:dsig_dy_Jpsi} we show our results for the cross section $d\sigma_{\rm incoh}/dy$ for incoherent $J/\psi$ production 
in Pb-Pb collisions. The dash-dotted lines show the results without the factor of eq. \ref{eq:K-factor}, while the dashed lines have 
the real-part/skewedness correction taken into account. We see, that the latter can give an enhancement of a factor $\sim 1.5$. 

The left panel shows the cross section for 
$\sqrt{s_{\rm NN}} = 2.76 \, \rm{TeV}$. We also show the data point taken by the ALICE collaboration \cite{Abbas:2013oua}. 
Our result with real-part/skewedness correction are in agreement with experiment. It appears that there is only little 
room left for diffractive production with nucleon dissociation. 

The nuclear suppression in our color-dipole Glauber-Gribov calculation
appears to be weaker than in the model of \cite{Guzey:2013jaa}, where inelastic shadowing corrections have been taken into account
in a different approach. The recent ``hot-spot'' models of \cite{Mantysaari:2017dwh,Cepila:2017nef} are also based on the color-dipole
approach, however they differ from us, and among each other, in the details of the nuclear averages. 
Some versions of these models give a similar nuclear suppression as the one obtained by us.

In the right panel of fig. \ref{fig:dsig_dy_Jpsi} we show our predictions for incoherent diffractive $J/\psi$ production 
in lead-lead collisions at $\sqrt{s_{\rm NN}} = 5.02 \, \rm{TeV}$.

Recently, the ALICE collaboration has observed a large enhancement of $J/\psi$ mesons carrying very small $p_T < 300 \, \rm {MeV}$
in the centrality classes corresponding to peripheral collisions \cite{Adam:2015gba}. 
This enhancement has been ascribed to coherent production \cite{Adam:2015gba,Klusek-Gawenda:2015hja,Zha:2017jch}.
It is interesting to estimate how large could be the contribution from incoherent diffraction discussed in this paper.
Notice that in transverse momentum the peaks of coherent and incoherent production are well separated. 
In fact for the incoherent $J/\psi$'s the maximum of the $p_T$-distribution is at $p_T \sim 300 \div 400 \, \rm{MeV}$, hence
the cut in \cite{Adam:2015gba} removes $\sim 35 \%$ of the incoherent contribution.

The theory of diffractive production in peripheral events, where nucleons from both nuclei undergo additional inelastic
interactions is not yet well developed. If we take at face value Contreras' determination\cite{Contreras:2016pkc} 
of photon fluxes  from the impact parameter region that corresponds to
the $70\div90 \%$ centrality class, we can estimate at $|y| = 3.25$, $\sqrt{s_{\rm NN}} = 2.76 \, \rm{TeV}$, $p_T^{\rm cut} = 300 \, \rm{MeV}$:

\begin{eqnarray}
{d\sigma_{\rm incoh} (AA \to V X|70\div90 \%) \over dy} &=&  n_{\gamma/A}(z_+|70\div90\%) \sigma_{\rm incoh}(W_+|p_T < p_T^{\rm cut}) \nonumber \\
 &+& n_{\gamma/A}(z_-|70\div 90\%) \sigma_{\rm incoh}(W_-| p_T< p_T^{\rm cut}) \nonumber \\ 
 &\approx& 15 \, \mu{\rm b} \, ,
\end{eqnarray}
which lies in the same ballpark as the estimate in \cite{Cepila:2017nef}.
The ALICE measurement \cite{Adam:2015gba} is 
\begin{eqnarray}
{d \sigma (AA \to VX | 70 \div 90 \%; 2.5 < |y| < 4.0 )\over dy} = 59 \pm 11 \pm 8 \,  \mu {\rm b} \, .
\end{eqnarray}
Within the present framework of ``peripheral photon fluxes'' 
incoherent $J/\psi$ alone can not explain the ALICE data in peripheral collisions, 
but can give a substantial contribution.

\section{Summary and outlook}

In this paper we have presented the Glauber-Gribov theory for incoherent photoproduction of vector mesons
on heavy nuclei within the color dipole approach. Here incoherent production means that the nucleus 
breaks up. There is a large rapidity gap between vector meson and nuclear fragments and no new particles
are produced in the nuclear fragmentation region.  
The color dipoles play the role of the eigenstates of the scattering
matrix and take into account the  inelastic shadowing corrections. We have developed the multiple scattering expansion
which involves matrix elements of the operator $\sigma^n(x,r) \exp[-\half \sigma(x,r)T_A(\bb)]$. We performed
calculations for $J/\psi$ and $\Upsilon$ photoproduction. Multiple scatterings lead to extended tails in 
the $t$-distributions. However due to the small dipole sizes relevant in these processes,
the multiple scattering terms are only important at large $t$, beyond the free-nucleon diffraction cone. 
The transverse momentum of heavy vector mesons in incoherent diffractive events will therefore
have the same  distribution as in coherent diffraction on a nucleon.

Our calculations are in agreement with data from ALICE in ultraperipheral lead-lead collisions 
at $\sqrt{s_{\rm NN}} = 2.76 \, \rm TeV$. There seems to be little room left for contributions
with nucleon dissociation. It will be interesting in the future to calculate also these contributions,
e.g. from the model presented in \cite{Cisek:2016kvr} in order to be able to predict the transverse
momentum spectra of vector mesons in the whole experimentally accessible phase space.

The nuclear excitation products contain neutrons which could be detected in forward 
neutron calorimeters. These neutrons are of different origin than those coming from
the electromagnetic excitation of giant dipole resonances, which are discussed e.g. in \cite{Pshenichnov:2011zz,Klusek-Gawenda:2013ema}.
Some ideas to exploit correlations of vector meson rapidity with neutron multiplicity from different mechanisms 
are discussed in \cite{Guzey:2013jaa}.

Incoherent diffractive production also contributes to the $J/\psi$ yield in peripheral inelastic heavy-ion collisions.
Rough estimates using photon fluxes of \cite{Contreras:2016pkc} give about $\sim 25 \%$ of the cross section
measured by ALICE \cite{Adam:2015gba}. 

Finally, our calculations can be extended to the deep inelastic region of interest for a possible future
electron-ion collider \cite{Caldwell:2010zza,Lappi:2014foa}. An extension to light vector mesons is
also straightforward, but may require modifications to the dipole cross sections used here which
we will explore elsewhere.

\section*{Acknowledgements}
We would like to thank Sasha Glazov (DESY) for help with the xFitter code and useful discussions. 
This work is supported by the Polish Ministry under
program Mobility Plus, no. 1320/MOB/IV/2015/0 and by the Polish National Science Center
grant DEC-2014/15/B/ST2/02528. 


\end{document}